\documentclass{imammb}

\jno{dqnxxx}

\usepackage{natbib}
\usepackage{graphicx,pst-xkey,xkeyval}
\usepackage[dvipsnames,svgnames]{pstricks}

\usepackage[english]{babel}
\usepackage{multirow}
\usepackage{amsthm,amssymb,amsfonts,amsmath,mathrsfs,color,xcolor,tikz,stackrel,subfigure}
\usepackage[colorlinks=true,urlcolor=blue,linkcolor=blue,citecolor=blue,anchorcolor=blue,filecolor=blue,menucolor=blue,pagecolor=blue]{hyperref}
\usepackage[version=3]{mhchem}
\DeclareGraphicsExtensions{.bmp,.png,.pdf,.jpg}
\usetikzlibrary{positioning,calc,snakes,patterns,shapes.misc,shapes.geometric,decorations.shapes,decorations}
\usetikzlibrary{calc,fadings,decorations.pathreplacing,3d}
\usepgflibrary{shapes.misc,arrows}

\usepackage{graphicx}
\usepackage{amsmath,amsthm,bm,mathrsfs}

\newcommand{\prt}{\partial}
\newcommand{\be}{\begin{equation}}

\newcommand{\ee}{\end{equation}}
\newcommand{\bd}{\begin{displaymath}}
\newcommand{\ed}{\end{displaymath}}
\newcommand{\eps}{\varepsilon}

\newcommand{\cd}{{\cal D}_{\vec{N}}}

\def\co1{c_{1}}
\def\cd2{c_{2}}
\def\ct3{c_{3}}

\definecolor{br}{rgb}{0.68,0.68,0.68}

\newcommand{\al}{\alpha}

\newcommand{\gm}{\gamma}

\newcommand{\dl}{\delta}

\newcommand{\Om}{\Omega}

\newcommand{\lm}{\lambda}

\newcommand{\del}{\nabla}

\numberwithin{equation}{section}

\begin{document}

\title{\bf A mathematical model for the progression of dental caries}
\author{ {\sc Rene Fabregas$^{1,2}$ and Jacob Rubinstein$^{1}$}\\[2pt]
$^1$Department of Mathematics, Technion-Israel Institute of Technology, \\
 Amado Building, Haifa, 32000, Israel.\\[6pt]
 $^2$Department of Applied Mathematics, Complutense University, \\
 S\'eneca 2 Street, Madrid, 28040, Spain.\\[6pt]
 }

\pagestyle{headings}
\markboth{Rene Fabregas and J. Rubinstein}{\rm A mathematical model for the progression of dental caries}
\maketitle


\begin{abstract}
{ A model for the progression of dental caries is derived. The analysis starts at the microscopic reaction and diffusion process. The local equations are averaged to derive a set of macroscopic equations. The global system includes features such as anisotropic diffusion and local changes in the geometry due to the enamel melting. The equations are then solved numerically. The simulations highlight the effect of anisotropy. In addition we draw conclusions on the progression rate of caries, and discuss them in light of a number of experiments.}
{Dental Caries, Mathematical Modeling, Numerical Simulation, Caries Progression Rate.}
\end{abstract}

\section{Introduction}

Dental caries is the process where the mineral components of the tooth dissolve and break down. The underlying process starts with bacteria in the mouth that digest foodstuff such a sucrose, and convert them into acidic material such as lactic acid. Hydrogen ions that dissolve from the lactic acid penetrate into the tooth and react with its mineral compounds such as enamel or dentin. As the enamel dissolves, the tooth become mechanically weak and susceptible to local collapse, which leads to the formation of cavities.

Our goal is to derive a mathematical model for the first part of this process, namely the dissolution of the enamel. The model starts from the local transport equations on the microscopic scale. The local equations are averaged via the method of homogenization to create a macroscopic model. Our model takes into account key factors such as the anisotropic geometry of the enamel and the time evolution of the microstructure due to demineralization and remineralization.

The outer layer of the tooth is formed of enamel. This is a mineral consisting of a packed array of rods, called
sometimes enamel prisms. A rod's length is about $8$ microns, and
its radius is about $1$ micron. In healthy enamel the rods are
separated by a distance under $1$ micron. Each rod
consists of hydroxyapatite crystals. Enamel is created in the
human body at infanthood. It is not generated later in life.
However, even upon demineralization through caries, the process is
reversible, and remineralization can take place.

In the tooth the rods are arranged roughly in layers, where the
long axis is perpendicular to the tooth surface. The inter-rod
domain consists of pores and of material that has the same
chemical composition as enamel, but the crystal orientation is not
ordered as in the rods themselves. It is known that diffusion in
the enamel is mostly along the prisms (rods).

Dentin is the material just below the enamel. It is not as hard as
enamel, but also not completely different from it. Dentin consists of
$70$ percent hydroxyapatite crystals, $20$ percent organic
material, and water. The mineral part of the dentin is not organized in rods
as in enamel. However, dentin also has a microscopic structure. It
has many tiny channels, dentinal tubules, containing fluid that
extend from the exterior part of it (the enamel-dentin border) to
the inner part of the tooth (pulp).

Caries is the process by which enamel is destructed.
This process starts when sugar consumed by the person (or animal) is digested by certain oral bacteria. The bacteria form a biofilm on the tooth surface.
Since the teeth are naturally abrased, for instance by tooth
brushing, these biofilms typically are created in protected areas
such as pits and grooves on the tooth surface or near the junction
of the tooth and the gums. The common terminology in the
literature is to call the area near the pit {\em occlusal
surface}, while the area at the lower part of the tooth is
sometimes termed {\em approximal surface}. The biofilm is
typically quite small and concentrated near the pit, while it is
spread when it forms at an approximal surface.

The bacteria convert the sugar into lactic acid. This acid reduces
the pH level, that is, increases the concentration of hydrogen ions,
near the tooth. The ions diffuse into the tooth, thus creating a high acidity
environment inside the tooth. This  leads to a reaction of the solid
(mineral) enamel according to
\begin{eqnarray}\label{req1}
\scalebox{1}{
\begin{tabular}{c}
\ce{$\underset{\text{Hydroxyapatite}}{\ce{Ca10(PO4)6(OH)2_{(s)}}}$ +8 H+_{(aq)} <=>[\quad \ce{k_1}\quad][\ce{k_{-1}}] 10Ca^2_{(aq)}+ +6HPO4^{2-}_{(aq)} +2H2O_{(l)}}
\\
\end{tabular}}
\end{eqnarray}
This reaction is reversible, but when the pH level drops to about
$5.5$ and lower it proceeds mostly to the right. As the enamel is reacting and is demineralized, material is lost and later cavities form in the enamel.

As the enamel dissolves, visible lesions can be observed. We shall focus here on the formation of such lesions where mineral is lost and the porosity increases. We do not consider
here the formation of cavities, which is a total mechanical
breakdown that occurs after too much mineral is lost.

The propagation of lesions was studied in the last two decades by
a number of techniques. We draw particular attention to the work of Bjorndal and
his colleagues \cite{bjo95}, \cite{bjo08}. The shape of the propagating lesion was summarized by Kidd and Fejerskov \cite{kidd}. When the lesion starts at an
approximal surface it has a triangular shape in the enamel, with a
base at the enamel surface, and a vertex at the enamel-dentin (ED) junction.
Another 'triangle' is then observed at the dentin, with a base at
the ED junction and a vertex deep in the dentin.
On the other hand, when the lesion starts from a biofilm at a pit
at the central top part of the tooth, it has the shape of a
triangle with a vertex at the pit (enamel surface) and a base at
the ED junction.

An interesting observation of the propagation of enamel lesions
and later on of dentin lesions is provided in \cite{bjo95}. Some
earlier theories  claim that the triangle base at the ED junction
forms after the lesion's tip reached the junction, and it is
caused by a lateral spread of the lesion at the junction. However,
Bjorndal et al. point out that the extent of the triangle base at
the ED junction is comparable to the extent of the biofilm
(triangular base of the lesion) at the enamel surface. From this,
and from other observations they conclude that the lesion base at
the dentin starts {\em before } the lesion reached the junction at
the enamel. Rather, they speculate that this base at the ED
junction is caused directly by the biofilm. This observation has
important clinical implications, as it implies that removing the
biofilm can arrest the lesion development.

We also point out this and other papers by Bjorndal and his
colleagues, according to which the bacteria hardly spreads into
the tooth before cavities are formed. Notice in particular
reference \cite{bjo08} for clinical significance of this theory.

In the next section we shall write down the microscopic equations. These equations will be homogenized in section 3, where we take into account the evolution of the microgeometry in space and time. The macroscopic equations involve averaged diffusion coefficients. These coefficients are determined by solving certain canonical differential equations. In section 4 we consider these equations and also discuss the difficult task of estimating the effective diffusivities in a varying local geometry. In section 5 we derive an evolution equation for the microstructure. A number of numerical examples are presented in section 6. Our model and suggestions for extending it are discussed in section 7. Finally, we derive in an appendix a few geometrical identities that we need for the homogenization calculations.

There is a long history of mathematical models for the initial stages of dental caries. For example, we refer to the work of \cite{zimmerman} and \cite{patela}. These models and others differ in the level of complexity (e.g. the number of compounds) and details of the reaction. They are one-dimensional and thus the anisotropy plays no role. Also, they are written down from the start on a macroscopic level. We start from the microscopic geometry, with the goal of deriving the macroscopic equations. In this process we highlight the importance of the anisotropy of the local enamel geometry, that implies anisotropic diffusion matrix for the different compounds. On the other hand, we shall replace the reaction term \eqref{req1} with a simpler model:
\begin{eqnarray}\label{req1b}
\scalebox{1}{
\begin{tabular}{c}
\ce{$\underset{\text{Hydroxyapatite}}{\ce{Ca10(PO4)6(OH)2_{(s)}}}$ +8 H+_{(aq)} <=>[\quad \ce{k_1}\quad][\ce{k_{-1}}] {\rm EMP}}
\end{tabular}}
\end{eqnarray}
where we use the notation EMP (enamel matrix products) to denote collectively the right hand side of equation \eqref{req1}. It is a simple matter to extend our model to account separately for the reaction and diffusion for each of the components of EMP.
\section{Local model}

We describe a simple model that incorporates some of the basic mechanisms
outlined above. The model is only for the formation of the lesion
and not formation of cavities. Moreover, in this paper we consider
only lesion formation in enamel.

We assume that the biofilm generates hydrogen ions at a given
rate. A typical time scale for this is an hour or two following food consumption, since the secretion of the acid takes place after meals.

We consider the enamel to occupy half space $(x_3>0, -\infty < x_1, \; x_2  < \infty)$.
The line $x_3=0$ is the tooth edge. We assume that the biofilm,
i.e. source of acid, occupies a smaller domain $x_3=0, -\dl < x_1,\;x_2
< \dl$. The concentration of $H^+$ ions is denoted by $H(x,t)$,
where $x=(x_1,x_2,x_3)$. There are several processes taking place in
the tooth. One is the diffusion and reaction of $H^+$ ions. A second is diffusion
and reaction of the various compounds on the right hand side of equation \eqref{req1}. For simplicity
we consider here only one such representative compound, denoted by EMP in equation \eqref{req1b}, and denote its concentration by $c(x,t)$.
The description above of the enamel surface and the underlying
reactions implies that the diffusion process is macroscopically
anisotropic. In fact, we expect caries to be controlled by two
general principles. One is that the diffusion coefficients depend
on the porosity of the solid, and this, in turn is affected by the
enamel density. The second is that diffusion is faster {\em
along} the rods than {\em across} them. To make the paper easier to read we describe in detail the diffusion and reaction of $c$, and then homogenize the equation for it. The equation for $H$ can be analyzed and averaged in a similar manner.

The local length scale for the underlying process will be denoted by us $l$. Because of the size of the enamel prisms and the typical distance between them we set $l=1 \; \mu$. To obtain proper nondimensional variables it is useful to write an ad-hoc one dimensional model (\cite{patela}, \cite{patelb}, \cite{anderson05}) consisting of a single
reaction diffusion system:
\begin{equation}
c_t = d c_{x_3 x_3} + R(c_s-c). \label{p1}
\end{equation}
Here $d$ is a diffusion coefficient and $R$ is a reaction rate. The parameter $c_s$ is the equilibrium value of the reaction, determined by the acidity, i.e. $H$, at any given point.

There is a great diversity of estimates for $d$ and $R$ in the literature. We choose here parameters following mostly the papers \cite{anderson05} and \cite{vanduk}. We thus set $d=10^4\;\mu^2 /h$, $R=10^{-2} \;1/h$, and $c_s=2 \cdot 10^{-3}\;g/cm^3$, which corresponds to acidity of about pH =4 \cite{anderson05}.

The values of $d$ and $R$ define naturally a length scale
\begin{equation}
\lm \sim \sqrt(d/R) = 100 \; \mu.\label{p3}
\end{equation}
This is the length scale where the diffusion and reaction are balanced. For example, consider equation \eqref{p1} in the semi infinite interval $x_3 >0$, with initial conditions $c(x_3,0)=c_s$. We assume that  $c$ approaches $c_s$ for large
values of as $x_3$. Assume further that the boundary condition at
$x_3=0$ is $c=c_s$, except for bursts of larger $c(0,t)$ of short
duration, resulting from a reduction in pH following a meal. In
these short bursts a reaction takes place according to equation
\eqref{p1}. The penetration length of the acidity is thus  $\lm$

We point out, though, that the diffusion coefficient $d$ is not fixed in time, since the melting of the enamel increases the size of the pores, and thus
increases $d$. Another point that should be taken into account is that the
reaction term (the second term on the right hand side of equation
\eqref{p1}) depends not just on $c$ but rather on the difference
$c_s-c$. When comparing the reaction term with the diffusion term
we assumed that the difference $c_s-c$ is of the same order as $c$
itself. However, when $c$ is near $c_s$, the reaction term is
practically smaller, and then the transition layer is longer. Finally we recall that $c_s$ is determined by the acidity as will be written explicitly below. We thus need to solve for both $c$ and $H$.

\section{Homogenization of the microscopic equations}

We assume that the enamel portion of the tooth consists of a
volume $E$ occupied by the rods and an inter-rod volume that we
denote by $V$. The ions and enamel matrix solute diffuse in the inter-rod domain $V$, and might react when they reach the rod boundaries $\prt E$. We
therefore write a diffusion equation for the concentration
$c(x_1,x_2,x_3,t)$  in $V$ and appropriate flux
condition at the  interface $\partial E$. Since the enamel melts
under the reaction with the $H^+$ ions, we must take into account
the time variation of $\prt E(t)$. We thus write the following
equations for the diffusion of $c$ and the reaction on $\prt E$:
\begin{eqnarray}
c_t = D\Delta c ,\;\;\; x \in V, \label{p5} \\ D\prt_n c =
R_0(c_s-c) -  \bar{v}_n c,\;\;\; x \in \prt E. \label{p7}
\end{eqnarray}

We use here $\bar{v}_n$ to denote the normal velocity of the rod
boundary $\prt E$. We scale the length by $\lm$: $x'=x/\lm$, and
time by $1/R$: $t' = R t$. In what follows we shall omit the
primes to simplify the notation.

Before writing down the scaled equations we describe our model for
the microstructure geometry. As noted above the local scale of the
prism, which we denoted by $l$, is much smaller than $\lm$.  We
thus define the small parameter $\eps = l/\lm$. We assume that $E$ consists of many
elongated prisms (rods) whose thickness is of $O(l)$. The prisms
main axis and also their cross section shape varies slowly on the
$\lm$ - scale. To distinguish between the two length scales we
introduce local variables $y_1=x_1/\eps,\; y_2=x_2/\eps$. Because
of the layered form of the prisms we assume that the geometry is
essentially periodic in the $y_1,y_2$ variables. However we do
allow for geometry variations also on the global $(x_1,x_2,x_3)$
scale since different parts of the enamel react at different
rates, determined by variations in $c$.

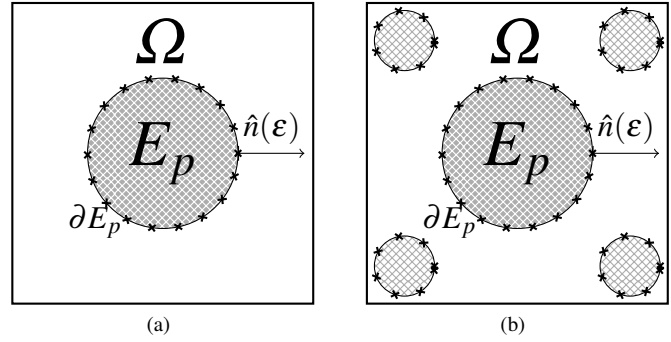
\begin{figure}[!h]
\centering
\subfigure[]{\label{cell1}
\begin{tabular}{c}
\begin{tikzpicture}[shorten >=.5pt,node distance=3.2cm,auto,decoration=crosses]
\draw[thick] (0,0) rectangle (4,4);
\def\mypath0{(2,2) circle (1cm)};
\fill [br] \mypath0;
\draw[pattern=crosshatch,pattern color=black!0] \mypath0;
\node [black,rotate=0,thick] at (2,2) {\Huge $E_p$};
\node [black,rotate=0,thick] at (2,3.5) {\Huge $\Omega$};
\draw [decorate,color=black, thick] (2,2) circle (1cm);
\draw [|->] (3,2) -- (3.9,2) node[midway,above] {\large $\hat{n}(\eps)$};
\node [black,rotate=0,thick] at (1.1,1.1) {\large $\partial E_p$};
\end{tikzpicture}
\end{tabular}
}
\subfigure[]{\label{cell2}
\begin{tabular}{c}
\begin{tikzpicture}[shorten >=.5pt,node distance=3.2cm,auto,decoration=crosses]
\draw[thick] (0,0) rectangle (4,4);
\def\mypath0{(2,2) circle (1cm)};
\draw[pattern=crosshatch,pattern color=br] (3.5,3.5) circle (0.4cm);
\draw[pattern=crosshatch,pattern color=br] (0.5,3.5) circle (0.4cm);
\draw[pattern=crosshatch,pattern color=br] (.5,0.5) circle (0.4cm);
\draw[pattern=crosshatch,pattern color=br] (3.5,0.5) circle (0.4cm);
\fill [br] \mypath0;
\draw[pattern=crosshatch,pattern color=black!0] \mypath0;
\node [black,rotate=0,thick] at (2,2) {\Huge $E_p$};
\node [black,rotate=0,thick] at (2,3.5) {\Huge $\Omega$};
\draw [decorate,color=black, thick] (2,2) circle (1cm);
\draw [decorate,color=black, thick] (3.5,3.5) circle (.4cm);
\draw [decorate,color=black, thick] (0.5,3.5) circle (.4cm);
\draw [decorate,color=black, thick] (.5,0.5) circle (.4cm);
\draw [decorate,color=black, thick] (3.5,0.5) circle (.4cm);
\draw [|->] (3,2) -- (3.9,2) node[midway,above] {\large $\hat{n}(\eps)$};
\node [black,rotate=0,thick] at (1.1,1.1) {\large $\partial E_p$};
\end{tikzpicture}
\end{tabular}
}
\caption{Examples for the periodic cell cross sections: (a)
Local periodic cell containing one prism, (b) Local periodic cell containing 5
prisms.}\label{gcell1}
\end{figure}

We denote the inter-rod domain in a single cell by $\Om$ and the
enamel cross section in a single cell is denoted by $E_p$. See
Figure~\ref{gcell1} for a sketch of simple two cases of a cells containing 1 prism (Figure~\ref{cell1}), and 5 prisms (Figure~\ref{cell2}) with circular cross sections. It is useful to express the assumptions above on the separation of scales by writing the boundary $\prt E_p$ of the enamel surface via an implicit function
\begin{equation}
\prt E_p : \{(x_1,x_2,x_3,y_1,y_2,t)|\;
F(x_1,x_2,x_3,y_1,y_2,t)=0\}. \label{h21a}
\end{equation}
We also need an expression for the normal vector $\hat{n}(\eps)$ to $\prt
E_p$ in terms of $F$. In fact, we need in the sequel
only the first two terms in the $\eps$ expansion of $\hat{n}$.
A straightforward calculation (see the Appendix) gives
\begin{equation}
\hat{n}(\eps) = \hat{\nu} + \eps\left( (\hat{\tau} \cdot
\nabla_2^x F)/G_F \; \hat{\tau}  +  \hat{k} F_{x_3}/G_F  \right),
\label{h21}
\end{equation}
where we introduced the notation $\hat{\nu}$ for the
two-dimensional normal to $\prt E_p$, i.e.
\begin{equation}
\hat{\nu}  = \nabla_2^y F/G_F,\;\;\; G_F=|\nabla_2^y F|.
\label{h21c}
\end{equation}
Here $\hat{\tau}$ denotes the unit two-dimensional tangent vector to
$\prt E_p$, and $\hat{k}$ is a unit vector in the $x_3$ direction.
Also, here and in what follows we use $\del^y_2$ and $\Delta^y_2$
to denote the gradient and Laplacian operators in the two
dimensional space $(y_1,y_2)$. Similarly, we use $\del^x_2$ and
$\Delta^x_2$ to denote the gradient and Laplacian operators in the
two dimensional space $(x_1,x_2)$.

The motion of the boundary of the prism $\prt E_p$ varies on the
$l$ local scale. Therefore we write
\begin{equation}
\bar{v}_n = R l v_n, \label{h23}
\end{equation}
so that $v_n$ is the nondimensional boundary normal velocity. To
complete the scaling we normalize the concentration $c$ by a representative value $\bar{c}$ of the equilibrium value $c_s$, i.e. $c'=c/\bar{c}$.

We are now at a position to write the nondimensional version of
equations \eqref{p5}-\eqref{p7}:
\begin{eqnarray}
c_t = \Delta c ,\;\;\; x \in V, \label{p9} \\ \prt_n c = \lm R_0/D
(c_s-c) - R l \lm/D v_n c ,\;\;\; x \in \prt E. \label{p11}
\end{eqnarray}

Because of the elongated shape of the rods, where the long axis
extends essentially in the $x_3$ direction, we expect the solution
to vary rapidly in the $x_1,x_2$ variables. Following the
convention in homogenization theory, we thus seek a
solution of the from $c(x_1,x_2,x_3,y_1,y_2,t)$. We assume that to leading order $c_s=c_s(x_1,x_2,x_3,t)$. Later on we shall justify this assumption. Since on the
present length and time scales we expect an averaged equation with
balanced diffusion and reaction, we also make the assumption that
$\lm R_0/D = \eps R_1$, with $R_1=O(1)$. Therefore, the boundary
condition \eqref{p11} becomes
\begin{equation}
\prt_{\nu} c = \eps\left(R_1(c_s-c) - v_{\nu}c - c_{x_3} F_{x_3}/G_F
- (\hat{\tau} \cdot \nabla_2^x F)/G_F \; \hat{\tau} \cdot
(\nabla_2^x c + 1/\eps \; \nabla_2^y c )  \right), \label{p11b}
\end{equation}
where $v_{\nu}$ is the two-dimensional velocity of the boundary. We use everywhere $\partial_{\nu}$ to denote the normal derivative in the $(x_1,x_2)$ plane, and
$\partial_{n}$ to denote the normal derivative in the $(x_1,x_2,x_3)$ space.

To solve for $c$ we expand
\begin{equation}
c(x_1,x_2,x_3,y_1,y_2,t) = c^0 + \eps c^1 + \eps^2 c^2 + ... \;\;,
c^i=c^i(x_1,x_2,x_3,y_1,y_2,t). \label{p13}
\end{equation}
Upon substituting the expansion into equation \eqref{p9} and the
boundary condition \eqref{p11b}, we obtain a cascade of equations
at different orders of $\eps$. At the leading order we obtain
\begin{eqnarray}
\Delta^y_2 c^0 =0 ,\;\;\; y \in \Om, \label{p15}\\ \hat{\nu} \cdot
\del^y_2 c^0 =0 ,\;\;\; y \in \prt E_p. \label{p17}
\end{eqnarray}
It is easy to verify that equations \eqref{p15}-\eqref{p17} imply
that $c^0$ is independent of the local scale $y$, i.e. $c^0 =
c^0(x_1,x_2,x_3,t)$.

At the next order we obtain the following equations for $c^1$:
\begin{eqnarray}
\Delta^y_2 c^1 =0 ,\;\;\; y \in \Om, \label{p19}\\ \hat{\nu} \cdot
\del^y_2 c^1 = -\hat{\nu} \cdot \del^x_2 c^0  ,\;\;\; y \in \prt
E_p. \label{p21}
\end{eqnarray}
We thus express $c^1$ in the form
\begin{equation}
c^1(x_1,x_2,x_3,y_1,y_2,t) = W^1(x_1,x_2,x_3,y_1,y_2,t) c^0_{x_1}
+ W^2(x_1,x_2,x_3,y_1,y_2,t) c^0_{x_2}. \label{p23}
\end{equation}
The {\em cell functions} $W^i$ solve a {\em cell problem} of the
form
\begin{eqnarray}
\Delta^y_2 W^i =0 ,\;\;\; y \in \Om, \label{p25}\\ \hat{\nu} \cdot
\del^y_2 W^i = -\hat{\nu}_i  ,\;\;\; y \in \prt E_p. \label{p27}
\end{eqnarray}
Notice that the cell problems are implicitly parameterized by
$(x_1,x_2,x_3,t)$ since $\prt E_p$ depends on these variables.

Proceeding to the next level in the expansion we find
\begin{eqnarray}
c^0_t = c^0_{x_3 x_3} + \Delta^x_2 c^0 + \Delta^y_2 c^2 + \del^y_2
\del^x_2 c^1 + \del^x_2 \del^y_2 c^1 ,
\;\;\; y \in \Om, \label{p29}\\
\nu \cdot \del^y_2 c^2 = -\hat{\nu} \cdot \nabla_2^x c^1 +
R_1(c_s-c^0) - v_{\nu}c^0 - c^0_{x_3} F_{x_3}/G_F - \nonumber \\
(\hat{\tau} \cdot \nabla_2^x F)/G_F \; \hat{\tau} \cdot
\left(\nabla_2^x c^0 + \nabla_2^y c^1\right), \;\;\; y \in \prt
E_p. \label{p31}
\end{eqnarray}
We integrate equation \eqref{p29} over the cell $\Om$, and write
\begin{equation}
|\Om| c^0_t = \int_{\Om} \nabla^x_2 \cdot (\nabla^y_2 c^1+
\nabla_2^x c^0)\;dy + \int_{\Om} \nabla^y_2 \cdot (\nabla^y_2 c^2+
\nabla_2^x c^1)\;dy=J_1 + J_2 . \label{p32a}
\end{equation}
The integral $J_2$ on the right hand side is transformed into a
boundary term, where we use the boundary condition \eqref{p31},
and the fact that $c^0$ does not depend upon $y$:
\begin{equation}
J_2 = |\prt E_p| R_1(c_s-c^0) - \left(\int_{\prt E_p} v_{\nu}\right)
c^0 - \left(\int_{\prt E_p} F_{x_3}/G_F \right) c^0_{x_3} -
\int_{\prt E_p} (\hat{\tau} \cdot \nabla_2^x F)/G_F \; \hat{\tau}
\cdot (\nabla_2^x c^0 + \nabla^y_2 c^1). \label{p32b}
\end{equation}
To compute the first integral $J_1$ in equation \eqref{p32a} we
use the following formula that we prove in the Appendix
\begin{equation}
J_1 = \nabla^x_2 \cdot \int_{\Om} (\nabla^y_2 c^1+ \nabla_2^x c^0)
\;dy + \int_{\prt E_p} \nabla^x_2 F /G_F \cdot (\nabla^y_2 c^1 +
\nabla_2^x c^0) \; dy. \label{p32c}
\end{equation}
The normal ($\hat{\nu}$) component of $(\nabla^y_2 c^1 + \nabla_2^x
c^0)$ vanishes on $\prt E_p$ in light of equation \eqref{p21}. The
tangential component of this term integrated in the second term on
the right hand side of equation \eqref{p32c} cancels with the last
term on the right hand side of equation \eqref{p32b}. Next we
define the $2 \times 2$ matrix
\begin{equation}
\bar{D}_{ij} = \int_{\Om} W^i_{y_j}\;dy. \label{p33}
\end{equation}
The computations of $J_1$ and $J_2$ and the definition \eqref{p33}
enable us to rewrite equation \eqref{p32a} in the form
\begin{equation}
|\Om|c^0_t = |\Om| c^0_{x_3 x_3} + \sum_{i=1}^2\prt_{x_i}
\left(\bar{D}_{ij} \prt_{x_j} \right)c^0 + |\prt E_p| R_1 (c_s-c^0) -
\left(\int_{\prt E_p} v_{\nu} \right) c^0 - \left(\int_{\prt E_p}
F_{x_3} /G_F \right) c^0_{x_3}. \label{p35}
\end{equation}
Equation \eqref{p35} is the averaged (homogenized) equation for
the evolution of $c^0$. To simplify it, and to recognize it as a
conservation law, we make use of two geometric identities that are
proved in the Appendix:
\begin{equation}
\int_{\prt E_p}v_n = |\Om|_t,\;\;\; \int_{\prt E_p} F_{x_3}/G_F =
-|\Om|_{x_3}. \label{p37}
\end{equation}

We now apply identities \eqref{p37} to rewrite equation
\eqref{p35} in the form
\begin{equation}
\left(|\Om|c^0\right)_t = \left(|\Om| c^0_{x_3}\right)_{x_3} +
\sum_{i=1}^2\prt_{x_i} \left(\bar{D}_{ij} \prt_{x_j}\right) c^0 +
|\prt E_p| R_1 (c_s-c^0). \label{p39}
\end{equation}
Notice that we wrote the derivatives of $c^0$ with respect to
$x_3$ and with respect to $(x_1,x_2)$ separately to emphasize the
symmetry breaking induced by the prism geometry.

Equation \eqref{p39} is not closed, since the enamel $E_p$ melts and
therefore both $|\prt E_p|$ and $|\Om|$ change in time and space.
In addition, the effective diffusion tensor $\bar{D}$ depends upon the concentration $c^0$. This issue will
be addressed in the next section. We still need to determine the solubility term $c_s$. From the simplified reaction equation \eqref{req1b} we have $c_t = k_1 X H - k_2c$, where we denote by $X$ the hydroxyapatite concentration. Therefore the term $c_s$ is proportional to the concentration $H$ of the hydrogen ions, and we write explicitly $c_s = \gm H$. Our estimate in the preceding section for $c_s$ at pH=4 implies $\gm \sim 40$.

We need also to evaluate the macroscopic equation for $H$. However, the hydrogen ions diffuse and react in the same geometry as the EMP. Therefore the same homogenization procedure that was performed in this section applies also to $H$. There are two differences, though. First, the hydrogen ions are much smaller than the EMP molecules, and therefore its diffusion coefficient is quite larger. Thus we assume that on the time scale on which $c$ evolves, the distribution of $H$ is at equilibrium. Still $H$ depends on time implicitly through the boundary conditions on the tooth surface, as will be explained in more detail via examples below. Secondly, the amount of ions lost to the reaction is small compared to the total amount of ions, and therefore we can neglect the reaction term. Thus, the equation for $H$ is
\begin{equation}
\left(|\Om| H_{x_3}\right)_{x_3} +
\sum_{i=1}^2\prt_{x_i} \left(\bar{D}_{ij} \prt_{x_j}\right) H =0.  \label{p41}
\end{equation}
We also write the final equation for the evolution of the EMP in the form
\begin{equation}
\left(|\Om|c\right)_t = \left(|\Om| c_{x_3}\right)_{x_3} +
\sum_{i=1}^2\prt_{x_i} \left(\bar{D}_{ij} \prt_{x_j}\right) c +
|\prt E_p| R_1 (\gm H -c), \label{p43}
\end{equation}
where we omitted the $0$ superscript for convenience. In section 6 we shall address the question of initial and boundary conditions for these two equations.

\section{The cell problem for the effective diffusion coefficients }

To solve the homogenized equations \eqref{p41} - \eqref{p43} we need to compute
the effective diffusion matrix $\bar{D}$. This requires solving the
cell problem \eqref{p25}-\eqref{p27}. Moreover, since the shape of
the enamel $E_p$ evolves in space and time, we need to solve the
cell problem for each time iteration and at all space points (on
the $x$ scale). To simplify this task, and since the local prim
changes slowly in space on its local $y$ scale, we assume that
the enamel motion is averaged, in the sense that the boundary
$\prt E_p$ moves normal to itself, and therefore the shape $E_p$ forms a
one-dimensional family of domains. The matrix $\bar{D}$ can be
computed for a number of representative shapes from this family, and
then the value of $\bar{D}$ for an actual shape can be determined by
interpolation from the computed family.

For example, we demonstrate this idea for the simple case where
the cell consists of the $2 \times 2$ square, and $E_p$ is a disc of radius
$\rho$ at its center, as depicted in Figure~\ref{cell1}. Symmetry implies that in this case $\bar{D}$ is of the form $\bar{D}  = dI$ where $d(\rho)$ is an effective diffusion
coefficient, and $I$ is the $2 \times 2$ identity matrix.
Solving the cell problem for this geometry gives the curve $d(\rho)$
as depicted in the dotted line in Figure~\ref{dff1}.

\section{Enamel dissolution}

In this section we couple the effective reaction diffusion equation to the
enamel dissolution process. The diffusion and reaction are described by equation \eqref{p43}. Recall that we used the length  $\lm$ as a length unit and
$1/R$ as a time unit. Also, the cell area $|\Om|$ and the prism boundary $|\prt E_p|$ are computed on the local $y$ scale ($l$), and therefore they are both $O(1)$ objects.

To verify that the model is consistent, and to estimate the parameter $R_1$, we give a rough estimate of the enamel dissolution. We thus concentrate on the global reaction term
\begin{equation}
\left(|\Omega| c \right)_t \sim  R_1 |\prt E_p| (c_s-c). \label{pa1}
\end{equation}
We return to dimensional units, using $1$h for the time variable and $1 \mu$ for the length variable. In light of the enamel basic symmetry we concentrate on a two dimensional slice of the enamel. The concentration $c$ is expressed in units of ${\rm mol}/\mu^2$. Indeed the relation  \eqref{pa1} is convenient for use in computing reaction terms since it expresses the chemical
quantities in units of moles. The reaction model \eqref{req1b}
relates the number of hydroxyapatite molecules to the number of EMP molecules. Denote the number of $c$
molecules in $\Om$ (per 1cm in the $x_3$ coordinate) by $N^c$, and similarly the
number of hydroxyapatite molecules in $E_p$ (per 1cm in the $x_3$ direction) is denoted $N^e$. We thus write
\begin{equation}
N^e_t = - N^c_t. \label{pa3}
\end{equation}

The reaction relation \eqref{pa1} can be written as
\begin{equation}
N^e_t \sim - R_1 |\prt E_p| (c_s-c). \label{pa5}
\end{equation}
To convert this result to the rate of change of $E_p$
itself we need to compute the molar concentration of
hydroxyapatite. This can be computed from the known value of the
molar mass of this molecule ($500 \; {\rm g/mol}$) and its known density
($3000 \;{\rm g/Liter}$). Thus the molar concentration of enamel is
\begin{equation}
\kappa^e = 6 \; {\rm mol/Liter} = 6 \cdot 10^{-11} \frac{{\rm mol}}{\mu^2 \;cm}. \label{pa11}
\end{equation}
Using the formula $\kappa^e = N^e/|E_p|$, and since $\kappa^e$ is
constant in time, we obtain
\begin{equation}
|E_p|_t = - \frac{|\prt E_p| R_1}{6} 10^{11} (c_s-c) . \label{pa13}
\end{equation}
For example, we assume that a cell is of size $2 \times 2$, and the prism radius is $1$. Thus there are approximately $0.25 \cdot 10^8$ prisms in a $1 \; {\rm cm}^2$ square. Assume further an acidity of pH=4, which corresponds to $c_s \sim 10^{-14} {\rm mol}/\mu^2$. Then, denoting the total enamel volume (per 1 cm in the $x_3$ direction) by $\tilde{E}$, we obtain $\tilde{E}_t \sim -2.5 R_1 \cdot 10^{-4} \;{\rm cm}^2/{\rm h}$. Recalling that the mass density of the enamel is about $3 \; {\rm g}/{\rm cm}^2$, we obtain that the rate of mass loss per unit area is
\begin{equation}
m_t \sim 7.5 R_1  10^{-4} {\rm g/h}\;{\rm cm}^2. \label{pa14}
\end{equation}
This compares favorably, using an $O(1)$ value for $R_1$, with experimental data \cite{holly}, \cite{anderson98}.

To complete the unknown parameters and functions in equation \eqref{p41} we need to obtain an equation for the volume $|\Om|$, the surface area $\prt E_p$, and the effective tensor $\bar{D}$. Our starting point is equation \eqref{pa13} which we write slightly differently as
\begin{equation}
|E_p|_t = - |\prt E_p| \frac{R_1}{\kappa} \left(c_s-c \right). \label{pb3}
\end{equation}
A precise melting equation requires an evolution equation not just for the area $|E_p|$ but also for the prisms shape. In general this involves a complex moving boundary problem. However, since the concentration $c$ varies slowly on the scale of individual prisms we propose a simplified model, as explained in the previous section. For example, we assume that initially $E_p$ is a disc, and this shape does not change appreciably as the prism melts. Thus the function $E_p(x,t)$ is characterized by a single parameter, namely the radius $\rho(x,t)$ of the prism. Therefore, we can express both the surface area $|\prt E_p|$ and the area $|\Om|$ using $\rho$. The evolution of $\rho$ is easily obtained form equation \eqref{pb3}. Denoting the cell, including the prism, by $\bar{\Om}$, we write $ |\Om| = |\bar{\Om}| -\pi \rho^2,\;\;  |\prt E_p| = 2\pi \rho$.
It follows that $\rho$ evolves according to
\begin{equation}
\rho_t = -\frac{R_1}{\kappa}  \left(c_s-c \right) = \frac{R_1}{6}\left(c_s-c\right), \label{pb3d}
\end{equation}
where in the last term we express the concentrations $c_s$ and $c$ in units of mol/Liter. Similarly, we can construct a more complex local prism geometry, such as a collection of discs of different radii at every cell, where each of the discs evolves by equation \eqref{pb3d}.

We now summarize the nonlinear diffusion-dissolution system that we derived: The concentration $c$ of the enamel matrix products is determined by equation \eqref{p43}. It is practically convenient to scale $c$ by $\gm$ and write $c=\gm q$, The scaled EMP concentration $q$ satisfies the equation
\begin{equation}
\left(|\Om|q\right)_t = \left(|\Om| q_{x_3}\right)_{x_3} +
\sum_{i=1}^2\prt_{x_i} \left(\bar{D}_{ij} \prt_{x_j}\right) q +
|\prt E_p| R_1 (H -q), \label{p43s}
\end{equation}
The area $|\Om|$ and surface area $|\prt E_p|$ are computed using equation \eqref{pb3d} for the local radius $\rho(x,t)$. for future reference we rewrite this equation in a scaled form:
\begin{equation}
\rho_t = -\frac{R_1}{6} \gm  \left(H-q \right). \label{pb3e}
\end{equation}
The ion concentration $H$ is in a quasi steady state, determined by equation \eqref{p41}. Finally, The diffusion tensor $\bar{D}$ is computed from equations \eqref{p25}-\eqref{p27} and the integral \eqref{p33}.

\section{Numerical examples}

In this section we consider the numerical implementation of the model derived in the previous sections. We also give numerical examples.

The evolution equations for $c$ (or $q$) and $H$ are solved in a finite domain.
Also, for simplicity we restrict the examples to two spatial dimensions. The anisotropy of the diffusion process is already manifested in this case. We thus consider in all the examples below the domain to be $$V=\{-1< x_1 <1,\;\; 0< x_3 <3\}.$$ This domain is similar to samples used in a number of experiments.

The diffusion equation \eqref{p43s} can be handled by standard methods for parabolic equations, such as Crank-Nicolson or ADI. One special feature is that we need to update at each time step the geometric functions $|\Om(x,t)| $ and $|\prt E_p(x,t)|$. As explained above, this can be done by updating at each time step the radius function $\rho(x,t)$. A second, more difficult point, is the computation of the effective diffusion tensor $\bar{D}$. Since the local cell geometry varies as a function of $x$ and $t$, we must solve the cell problem \eqref{p25}-\eqref{p27} and then compute the integral \eqref{p33} for each local value of $\rho(x,t)$. For instance, when solving the equations numerically, we need to find this tensor at each grid point (in space and time). This is a hard problem even under the simplification that $E_p$ is characterized by a single disc. We thus propose a method to overcome this problem. The idea is to compute {\em apriori} the tensor $\bar{D}$ as a function of the radius $\rho$ for a finite set of values $(\rho_1,\rho_2,..,\rho_n)$. Then, at each point $(x,t)$ we interpolate $\bar{D}(\rho(x,t))$ from the precomputed set of $\bar{D}$.

\begin{figure}[!h]
\centering
\subfigure[]{\label{dff1}
\includegraphics[width=6cm]{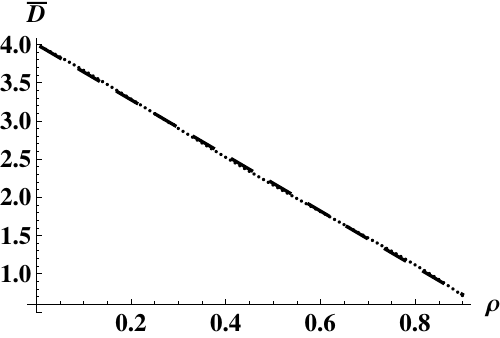}
}
\subfigure[]{\label{dff2}
\includegraphics[width=6cm]{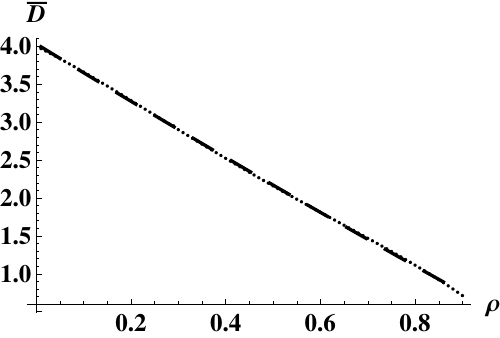}
}
\caption{The effective diffusion coefficient $\bar{D}(\rho)$
for the geometry of Figure~\ref{cell1} as a function of the local geometry (dotted line)
compared with approximations of it (dashed lines). (a) Linear interpolation $\bar{D}(\rho)=4 - 3.64 \rho$, (b) Quadratic interpolation $\bar{D}(\rho)=4.03 - 3.82\rho + 0.2\rho^2$.}\label{dffg1}
\end{figure}

We examined several interpolation models. For example, consider the case where each cell has a single disc. In Figure~\ref{dff1} we compare the exact $\bar{D}$ for this case to a linear interpolation of it, and in Figure~\ref{dff2} we compare a quadratic interpolation to $\bar{D}$ with the exact value. It seems that the quadratic interpolation is essentially the same as the exact value; however, since our geometry itself is only a model, even the linear interpolation is very good. A similar interpolation can also be shown for other local geometries, such as the one depicted in Figure~\ref{cell2}.

\vskip 0.5cm \noindent {\bf Example 1:} Our first test is carried out for the reaction diffusion alone, neglecting the effect of changes in the enamel geometry. This is certainly justified for short time periods, following the estimate we gave in the previous section on the dissolution rate for pH value of about 4. The cell consists of a single disc of radius $\rho$. Symmetry implies that in this case the effective diffusion tensor $\bar{D}$ is diagonal, and therefore the suppression of the $x_2$ variable does not have a serious effect on the problem. For this geometry we computed the effective horizontal diffusion coefficient for $\rho = 0.9$ and found  $\bar{D}=0.72$.

We solve for the specific scenario where the system is at equilibrium, and then is disturbed by a burst of acidity (low pH) for a certain time period.
To fix ideas, we assume that $H=10^{-5}$ at all the boundaries except at the top of the tooth ($x_3=0$ in our convention). For the time period, $0 < t < T=72$ we simulate the burst of low pH by setting the boundary condition $H(x_1,0;t)= 10^{-5}+9\cdot10^{-5}A(0.1-|x_1|)$, where $A(\xi)$ is the Heaviside function.

Given $H$, and neglecting the change in $\Om$ and $|\prt E_p|$ we solve equation \eqref{p43s}. We set the boundary condition $q=10^{-5}$ over the entire boundary. While this is a reasonable condition on the lateral boundaries, and also deep inside the tooth, it is not so obvious at the tooth edge $x_3=0$. However, we point out that this condition stabilizes $q$ in the sense that it acts to inhibit demineralization. This is consistent with the physiology of the tooth, where there is an external thin layer near the surface that is resilient to demineralization. The initial condition is set to be $q \equiv 10^{-5}$.

Because of the similarity of the equations and boundary conditions for $H$ and $q$, we can simplify the calculations by defining the difference function $v(x,t)= H-q$. Thus, $v$ solves
\begin{equation}
\left(|\Om|v\right)_t = \left(|\Om| v_{x_3}\right)_{x_3} +
\sum_{i=1}^2\prt_{x_i} \left(\bar{D}_{ij} \prt_{x_j}\right) v -
|\prt E_p| R_1 v, \label{p43t}
\end{equation}
with the appropriate boundary conditions set above.

Upon solving for $q$ and $H$ (or for $v$), we can compute the evolution of $\rho(x,t)$ according to equation \eqref{pb3e}. Recall that in this example, while we compute $\rho(x,t)$ we do not change the geometry. Since the change in $\rho$ is small in this time interval, we express $\rho$ in the form $\rho=\rho_0 +10^{-4}\rho_1(x,t)$, where $\rho_0$ is the original (constant) value of $\rho$. Therefore $\rho_1(x,t)$ solves the equation
\begin{equation}
\prt_t \rho_1= -\frac{10^4R_1}{6}\gm (H-q) = -\frac{10^4R_1}{6}\gm v, \label{pb3f}
\end{equation}
where  $R_1=1$.

\begin{figure}[!h]
\centering
\subfigure[]{\label{fv11}
\includegraphics[width=6cm]{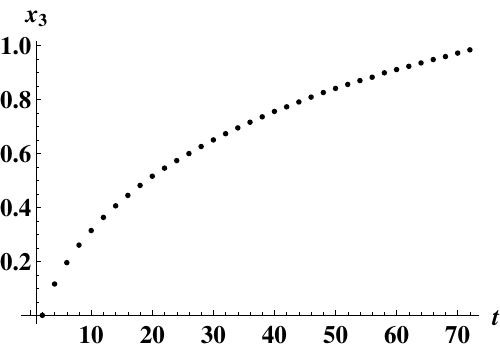}
}
\subfigure[]{\label{fv12}
\includegraphics[width=6cm]{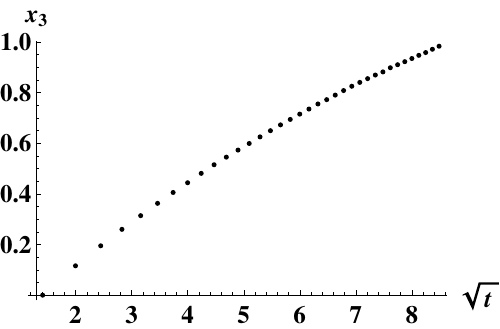}
}
\subfigure[]{\label{fv13}
\includegraphics[width=6cm]{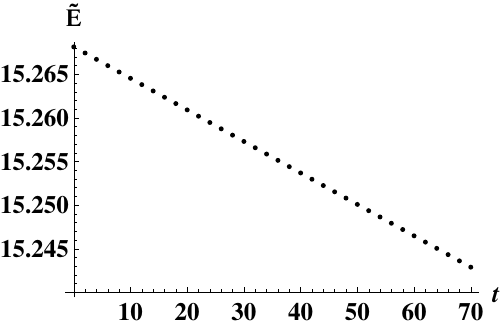}
}
\caption{Progression of caries for the first boundary condition
of Example 1: (a) Motion of $x_3$, the tip of a level set of $\rho_1$, (b)
Motion of the same level set as a function of $t^{1/2}$, (c) The evolution of
the total enamel volume $\tilde{E}(t)$}\label{fv10}
\end{figure}

The progression of caries, as calculated in this example, is presented in Figure~\ref{fv10}. In Figure~\ref{fv11} we draw the tip of the level set $\rho_1=-15$ as a function of $t$. In Figure~\ref{fv12} we draw the same level set tip, but now as a function of $t^{1/2}$. In Figure~\ref{fv13} we depict the total enamel volume $\tilde{E}$ as a function of $t$. We chose these specific graphs in order to compare our model, at least quantitatively, with well-known experiments. There is a debate in the literature as to what is the actual rate of caries progression. Anderson et al. \cite{anderson98} cite a number of authors that argue that caries progresses as a linear function of $t^{1/2}$.  A typical result along this line is the paper of Poole et al. \cite{poole}. On the other hand, many authors, and in particular Anderson himself, provide experimental evidence that caries progresses linearly in $t$.

We argue that the progression rate depends on how one defines it. This can be seen from the present example. When the total enamel volume lost to reaction is computed, which corresponds to the decline in $\tilde{E}$ in our case, the progression is linear. This follows directly from our equations. Consider first the evolution of $\rho$ \eqref{pb3e}. The evolution of $|\tilde{E}|_t$ is proportional to the integral of $\rho \rho_t$. From equation \eqref{pb3e} this, in turn, is proportional to the integral of $\rho v$. Since the evolution of $v$, which is determined by equation \eqref{p43t}, stabilizes in a few hours, the integral of $\rho v$ is proportional to the total flux of $v$ (a conclusion that is simply the law of mass conservation). However, since the evolution of $\rho$ is slow on the relevant time scale, the flux is approximately constant in time, and therefore the evolution of $\tilde{E}$ is linear in time. This argument is indeed manifested in Figure~\ref{fv13}.

A number of authors, including for example \cite{poole} and  \cite{featherstone}, claim that the progression of caries is proportional to $t^{1/2}$. The theoretical justification, see for example \cite{vanduk}, is that for a short time period caries is dominated by diffusion, and at least initially, and in a one-dimensional approximation, the concentration of the ions is proportional to $\mbox{Erf}(x_3/2(Dt)^{1/2})$. However, with the observed value of the diffusion coefficient $D$, and the typical scaling for caries, this description is only valid for a couple of hours at most. Therefore it cannot be used for experiments like those in reference \cite{poole}, for example, where the enamel was exposed to large acidity for days.

\begin{figure}[!h]
\centering
\subfigure[]{\label{fv21}
\includegraphics[width=6cm]{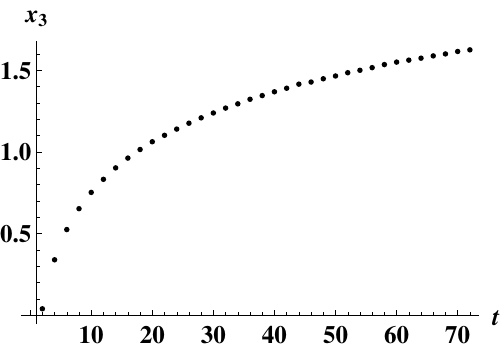}
}
\subfigure[]{\label{fv22}
\includegraphics[width=6cm]{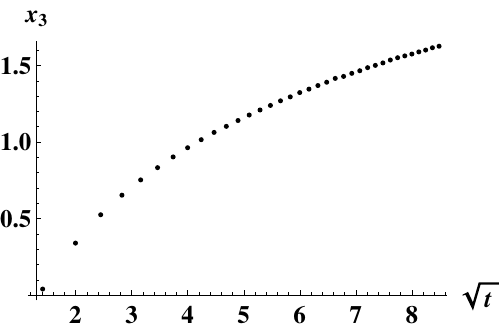}
}
\subfigure[]{\label{fv23}
\includegraphics[width=6cm]{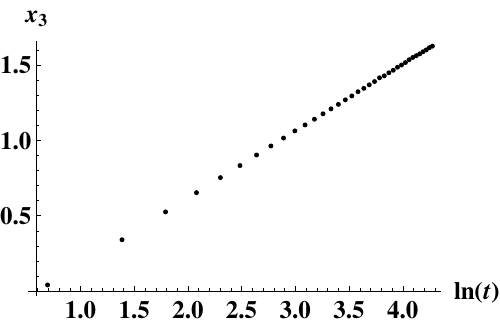}
}
\subfigure[]{\label{fv24}
\includegraphics[width=6cm]{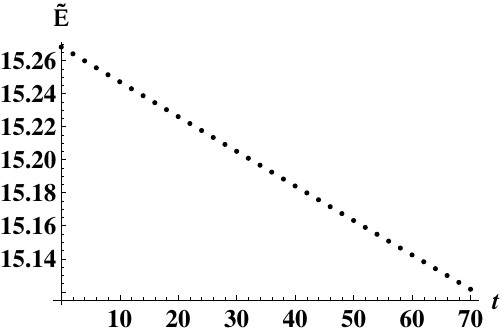}
}
\caption{Progression of caries for the boundary condition $v(x_1,0,t)=10^{-4}\cos (\pi x_1/2)$: (a) Motion of $x_3$, the tip of a
level set of $\rho_1$, (b) Motion of the same level set as a function of $t^{1/2}$, (c)
Motion of the same level set as a function of $\ln t$, (d) The evolution of the
total enamel volume $\tilde{E}(t)$.}\label{fv20}
\end{figure}

The authors in references \cite{poole} and \cite{featherstone} (and similar publications) visually inspected  the lesion, and deduced the progression from this inspection. This is similar to following the tip of a specific value of $\rho$, as depicted in Figure~\ref{fv11}. It is tempting to associate this curve with some power law. For instance, we drew in Figure~\ref{fv12} the motion of the level set as a function of $t^{1/2}$. This curve looks approximately linear. However, selecting a power law can be misleading. We draw attention to the recent analysis of Stumpf and Porter \cite{stumpf}. They showed that power law expressions empirical studies is often unjustified. To demonstrate this point in the present context, we performed another numerical test, where we replaced the boundary condition above for $H$ and $q$ by the boundary condition (for their difference)
$v(x_1,0,t) = 10^{-4} \cos (\pi x_1/2)$.

The advantage of selecting this boundary condition is that after a short time period the function $v$ reaches a steady state that can be solved explicitly. For the specific parameters in this example we have that in the steady state $v$ solves $v_{x_3 x_3} + 1/2 v_{x_1 x_1 } - 3.87 v=0$. Thus, approximately we have
$$v \sim 10^{-4} \cos(\pi x_1/2) e^{-2.25 x_3}.$$ The equation for $\rho$ then implies that level sets of $\rho$ evolve (for $t > 1$, say) approximately according to
\begin{equation}
x_3(t) = 0.44 \ln t + C, \label{p51}
\end{equation}
for some constant $C$. Indeed the computed curve of $x_3(t)$ is shown in Figure~\ref{fv21}, and the linear dependency of $x_3$ on $\ln t$ is clearly observed in Figure~\ref{fv23}, with the predicted slope.

\vskip 0.5cm \noindent {\bf Example 2:} We compare here two  simulations. We use the same geometry as in Example 1, with boundary condition $q=10^{-5}$ over the entire boundary and initial condition $q \equiv 10^{-5}$. Also, we assume that $H=10^{-5}$ at all the boundaries except at the top of the tooth, but. We solve for the time period, $0 < t < T=264$ (11 days), and the boundary conditions at $x_3=0$ are given by $H(x_1,0;t)= 10^{-5}+5\cdot10^{-4}A(0.1-|x|)$, which simulates relatively low pH.

\begin{figure}[!h]
\centering
\subfigure[$t=264$]{\label{f264a}
\includegraphics[width=3.35cm]{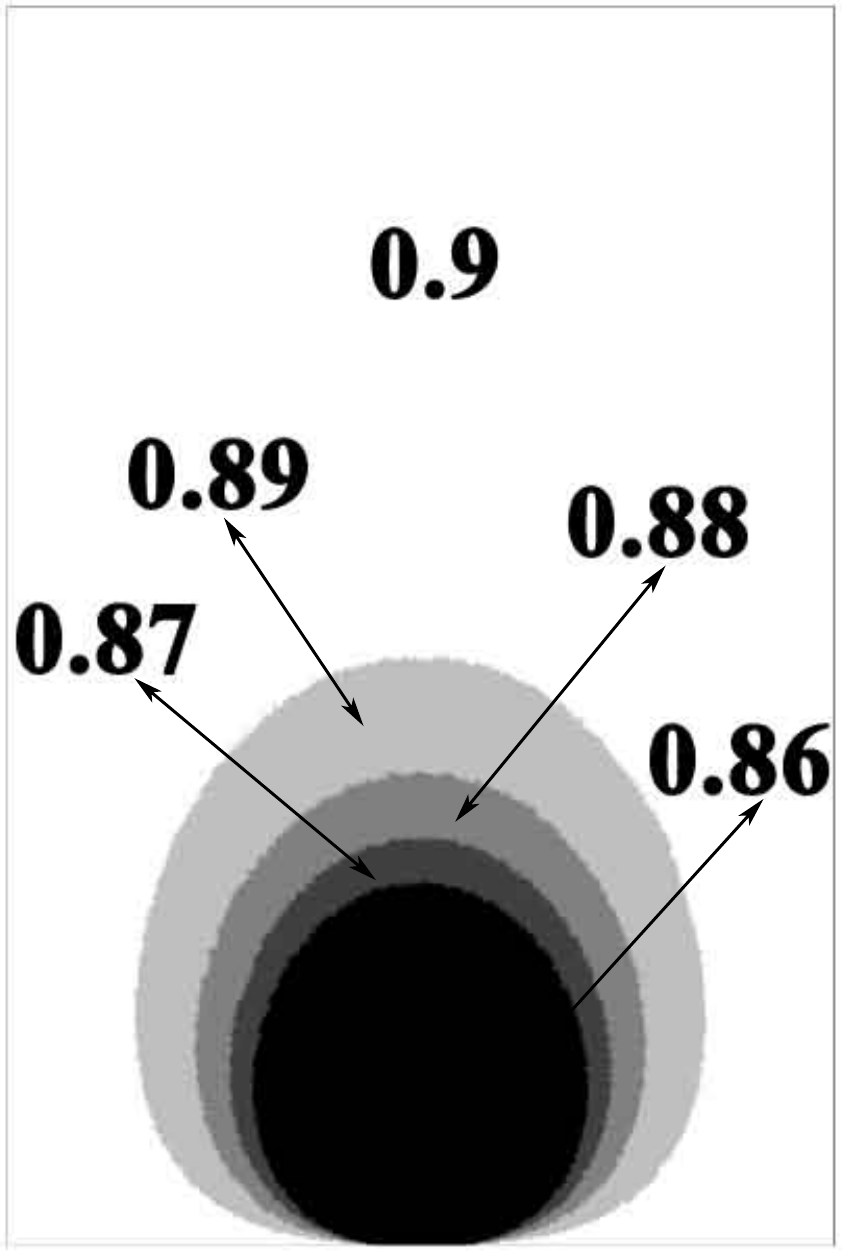}
}
\subfigure[$t=120$]{\label{f120}
\includegraphics[width=3.35cm]{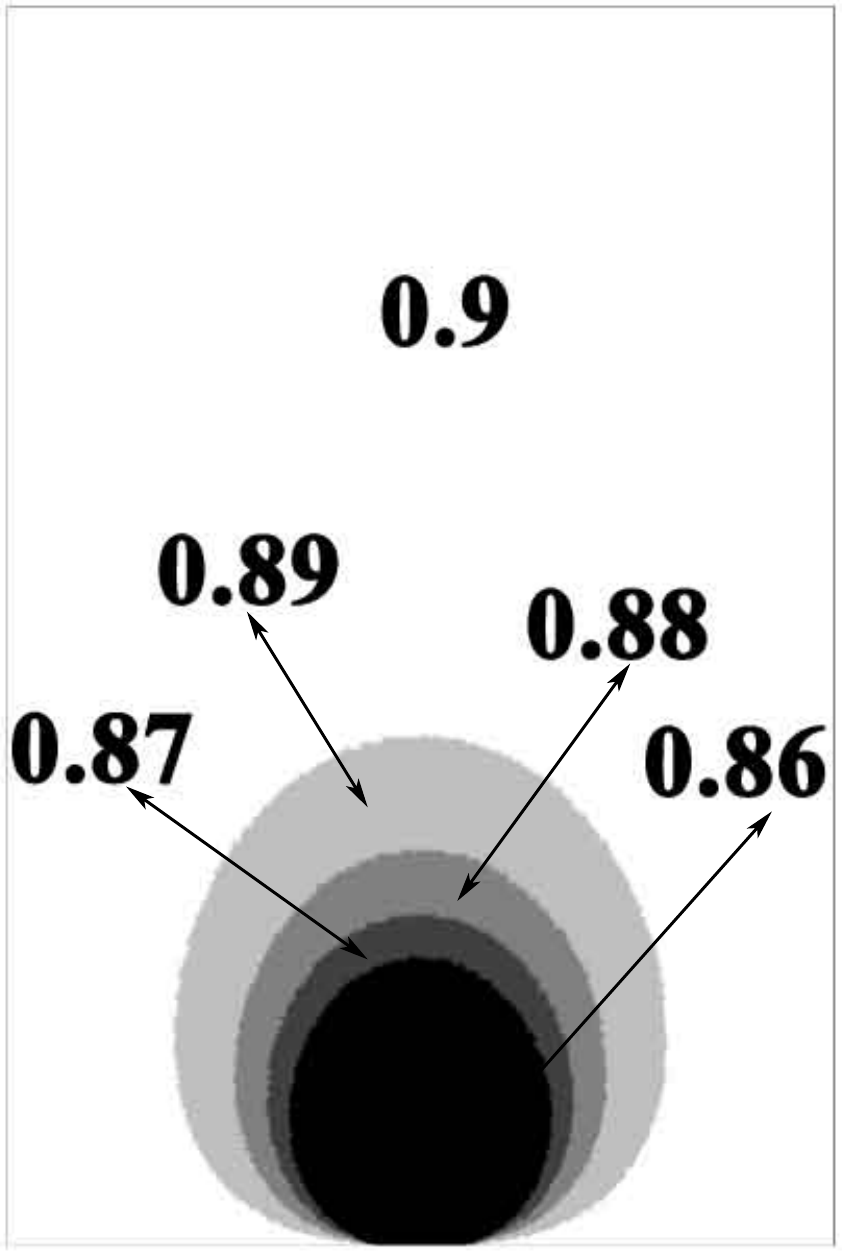}
}
\subfigure[$t=192$]{\label{f192}
\includegraphics[width=3.35cm]{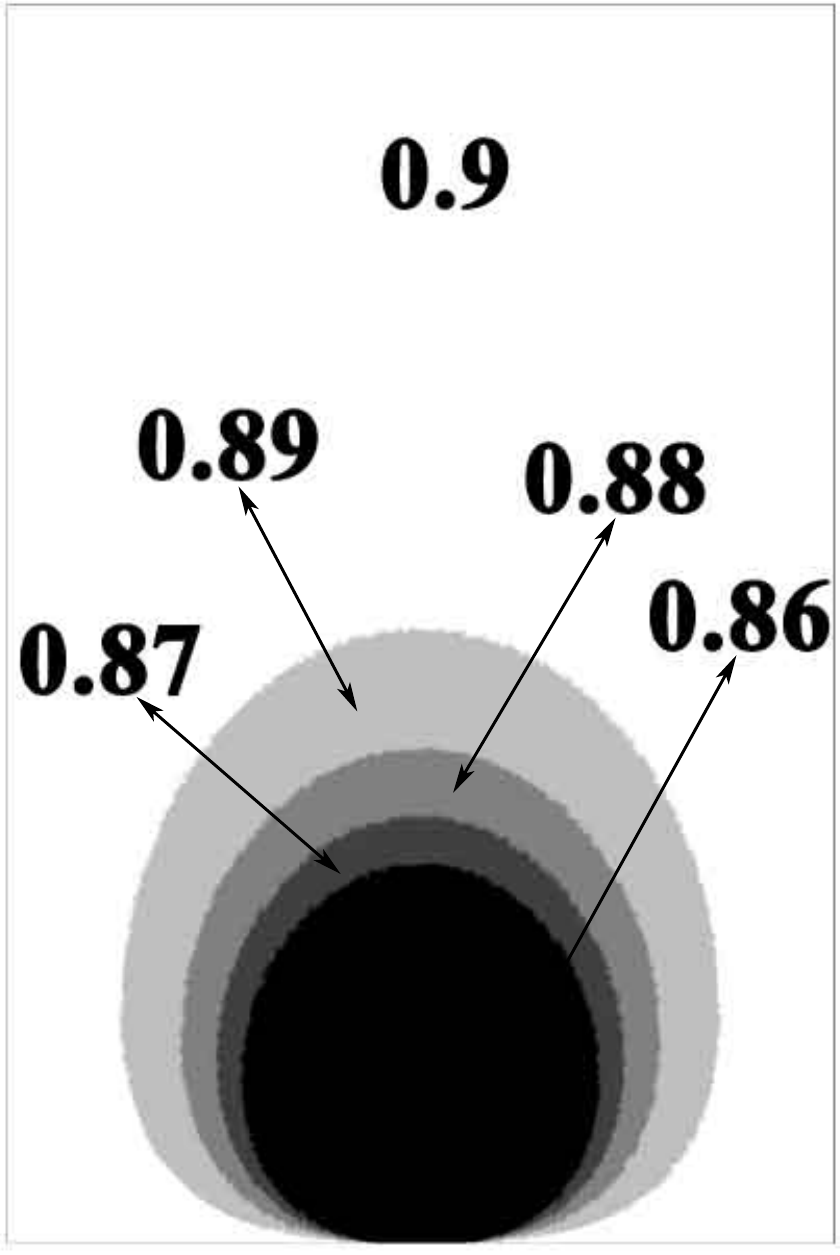}
}
\subfigure[$t=264$]{\label{f264b}
\includegraphics[width=3.35cm]{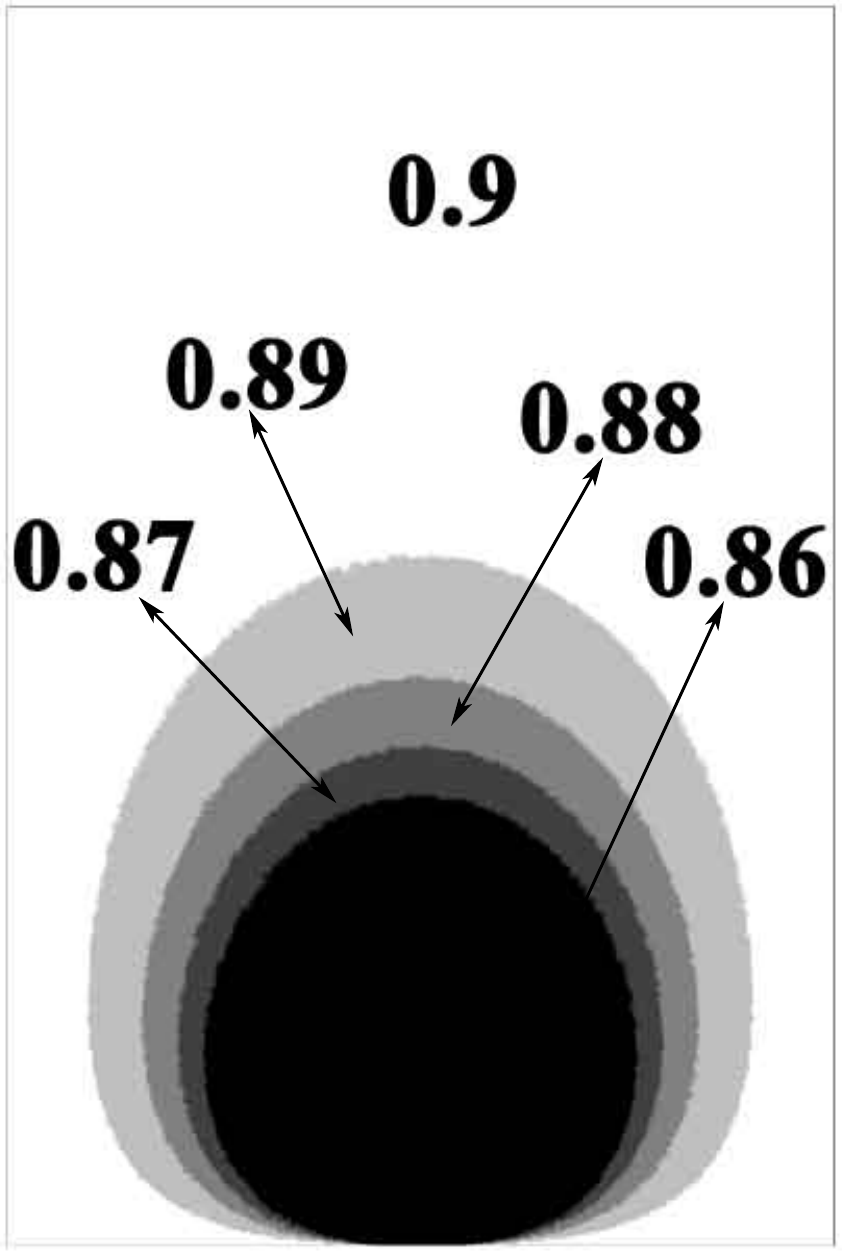}
}
\caption{(a) Level sets of $\rho(x,t=264)$ for the boundary
conditions of Example 2, neglecting the evolution of the geometry and of
$\bar{D}$. (b)-(d) Level sets of $\rho(x,t)$ for three values of $t$ for the
boundary conditions of Example 2. Here the evolution of the geometry and of
$\bar{D}$ was taken into account.}\label{exr1}
\end{figure}

In the first simulation (Figure~\ref{f264a}) we fixed the geometry, i.e. $\rho$ at its initial value $\rho=0.9$. we compute $\rho(x,t)$ as in Example 1 above, and draw the level sets of $\rho(x,264)$. In the second simulation we update $\rho(x,t)$ throughout the evolution of $v$. We draw for comparison the level set at $\rho(x,120)$, (Figure~\ref{f120}), $\rho(x,192)$ (Figure~\ref{f192}), and $\rho(x,264)$ (Figure~\ref{f264b}). Since the enamel melts appreciably in this long time interval, $\rho$ decreases and thus $\bar{D}$, which is now a function of space and time increases. As a results, we see a faster progression of caries in Figure~\ref{f264b} than in Figure~\ref{f264a}.

\section{Discussion}

We presented a model for the progression of dental caries lesions. We start from a model for the microscopic geometry. The local reaction diffusion equations are homogenized, while taking into account the change of the geometry in space and time because of the dissolution of the enamel. The global macroscopic equations include important features such as anisotropic diffusion, resulting from the local geometry, an effective reaction term, and an equation for the evolution of the geometry. The equations are nonlinear due to the mutual coupling of the reaction-diffusion and the geometry.

To demonstrate the effective equations we considered the progression of caries in a rectangular domain. We chose this geometry for its simplicity and because it is essentially the geometry used in many experiments. In our examples we concentrated on two issues. One is the effect of the anisotropy and of the melting. This is best seen in Example 2 in the preceding section. Caries is seen to propagate faster along the "easy" $x_3$ axis, than in the orthogonal direction. The effect of dissolution is observed by comparing Figure~\ref{f264a} to Figure~\ref{f264b}. In both cases the equations were solved for a long period of time under conditions of low pH. However, in the first case we neglected the change in the geometry, and in the second case we updated the geometry, and also the diffusion coefficient. Indeed in the second case caries propagated faster also in the horizontal direction.

Another goal of our numerical simulation was to consider the question of progression rate of caries. A number of conflicting theories appear in the literature. We demonstrated in the preceding section that the progression rate depends on the way it is defined. We also pointed out that some of the power laws proposed in the literature are questionable.

The preliminary model presented here can be extended in a number of ways. First, the model can be embedded in a realistic tooth geometry. This will enable to study the effect of the tooth surface on caries progression. Another extension, that we already pointed out in the introduction is to consider separate reaction diffusion equation for the different compounds in the reaction. A third extension is to build into the model the hard external layer near the tooth surface that is more resilient to caries. Furthermore, adding compounds such as fluoride to the model will enable applying the model to study medical treatments for arresting caries. We are indeed working on upgrading our model in these directions.

\section{Appendix: Proof of some geometric formula}

We prove in this appendix some of the geometric formula we used earlier in deriving
the averaged equation of the concentration $c$.

To prove formula \eqref{h21} for the normal $\hat{n}$ we write
$$\hat{n} = \alpha\left(F_{y_1}+\eps F_{x_1},F_{y_2}+\eps F_{x_2}, \eps F_{x_3}\right),$$ where $\alpha$ is a factor chosen so $\hat{n}$ has unit norm. To leading order we obtain $$ \al = 1/G_F\left(1-\eps \del^y_2F \cdot \del^x_2 F/G_F^2\right).$$
Therefore $$\hat{n} = \hat{\nu} +\eps/G_F \left(\del^x_2F - \hat{\nu}(\hat{\nu} \cdot \del^x_2 F) +
\hat{k} F_{x_3}\right).$$ Formula \eqref{h21} follows from the last identity.

To prove the first formula of \eqref{p37} we observe that the volume of $\Om$ is $|\Om| = 1- |E_p|$, can also
be written as
as $$|\Om| = (1-\int_{cell} \chi_{E_p}\;dy) ,$$ where $\chi_{E_p}$
is the characteristic function of $E_p$. We then notice the
relation $\prt_t \chi_{E_p} = \dl(x-\prt E_p) v_{\nu}$.
Substituting the last relation into the equation
\begin{equation}
\prt_t |\Om| = -\int_{cell} \prt_t \chi_{E_p}\;dy \label{app1}
\end{equation}
leads to the first formula of \eqref{p37}.

The second identity of equation \eqref{p37} and identity \eqref{p32c} follow from the following general argument: Let
$\theta$ be any of the arguments $(x_1,x_2,x_3)$ of the function $F$ introduced in equation \eqref{h21a}.
To emphasize that we concentrate on this argument we write $$\prt E_p(\theta): \{F(y_1,y_2,\theta)=0\}.$$ We now show that
\begin{equation}
\frac{\prt}{\prt \theta} |E_p| = -\int_{\prt E_p} F_{\theta}/G_F. \label{app21}
\end{equation}

To find the change in the area of $E_p$ as we vary the parameter $\theta$ we need to find the area of ring
between $E_p(\theta)$ and $E_p(\theta+ d\theta)$.  To find the thickness of the ring we move a distance $\delta$ along the normal $\nu$. To determine $\delta$  we expand the relation $F(y_1 + \dl F_{y_1}/G_F, F_{y_2} + \dl F_{y_2}/G_F, \theta
+\delta\theta)=0$. Using also $F(y_1,y_2,\theta)=0$ we obtain $\dl = -F_{\theta} d\theta/G_F$. Therefore
$d \dl/d\theta = -F_{\theta}/G_F$. This proves \eqref{app1}

\section*{Acknowledgment}

This work is supported in part by a F.I.R.S.T. Initial Training Grant of the
ERC, and in part by a grant from the ISF. We thank Dr. Marc Campillo for useful
discussion on caries and for bringing a few references to our attention.

\vspace*{6pt}


\end{document}